\documentclass[traditabstract]{aa}

\usepackage{epsfig,graphicx,natbib}
\usepackage{longtable}
\usepackage{amssymb}
\usepackage{nicefrac}

\def\Tcmb{\hbox{$T_\mathrm{CMB}$}}

\def\Tex{\hbox{$T_\mathrm{ex}$}}
\def\Tkin{\hbox{$T_\mathrm{kin}$}}

\def\nH2{\hbox{$n_\mathrm{H_2}$}}
\def\kms{\hbox{km\,s$^{-1}$}}
\def\PKS1830{\hbox{PKS\,1830$-$211}}
\def\percm{\hbox{$\rm cm^{-2}$}}
\def\percm3{\hbox{$\rm cm^{-3}$}}
\def\fH2{\hbox{$f_{\rm H2}$}}

\def\fc{\hbox{$f_c$}}

\begin{document}

\title{Detection of deuterated molecules, but not of lithium hydride, in the z=0.89 absorber toward \PKS1830 \thanks{Spectra are available in electronic form
at the CDS via anonymous ftp to cdsarc.u-strasbg.fr (130.79.128.5)
or via http://cdsweb.u-strasbg.fr/cgi-bin/qcat?J/A+A/}
}

\author{S.~Muller \inst{1}
\and E.~Roueff \inst{2}
\and J.~H. Black \inst{1}
\and M.~G\'erin \inst{3}
\and M.~Gu\'elin \inst{4}
\and K.~M.~Menten \inst{5}
\and C.~Henkel \inst{5,6}
\and S.~Aalto  \inst{1}
\and F.~Combes \inst{7}
\and S.~Mart\'in \inst{8,9}
\and I.~Mart\'i-Vidal \inst{10,11}
}
\institute{Department of Space, Earth and Environment, Chalmers University of Technology, Onsala Space Observatory, SE-43992 Onsala, Sweden
\and LERMA, Observatoire de Paris, PSL Research University, CNRS, Sorbonne Universit\'e, 92190 Meudon, France
\and LERMA, Observatoire de Paris, PSL Research University, CNRS, Sorbonne Universit\'e, 75014 Paris, France
\and Institut de Radioastronomie Millim\'etrique, 300, rue de la piscine, 38406 St Martin d'H\`eres, France 
\and Max-Planck-Institut f\"ur Radioastonomie, Auf dem H\"ugel 69, D-53121 Bonn, Germany
\and Astron. Dept., King Abdulaziz University, P.O. Box 80203, Jeddah 21589, Saudi Arabia
\and Observatoire de Paris, LERMA, College de France, CNRS, PSL Univ., Sorbonne Univ., F-75014, Paris, France
\and European Southern Observatory, Alonso de C\'ordova, 3107, Vitacura, Santiago 763-0355, Chile 
\and Joint ALMA Observatory, Alonso de C\'ordova, 3107, Vitacura, Santiago 763-0355, Chile
\and Observatori Astron\`omic, Universitat de Val\`encia, Parc Cient\'ific, C. Catedr\`atico Jos\'e Beltr\'an 2, 46980 Paterna, Val\`encia, Spain
\and Departament d'Astronomia i Astrof\'isica, Universitat de Val\`encia, C. Dr. Moliner 50, 46100 Burjassot ,Val\`encia, Spain
}

\date {Received  / Accepted}

\titlerunning{D-species toward \PKS1830}
\authorrunning{S. Muller et al.}

\abstract{Deuterium and lithium are light elements of high cosmological and astrophysical importance. In this work we report the first detection of deuterated molecules and a search for lithium hydride, $^7$LiH, at redshift z=0.89 in the spiral galaxy intercepting the line of sight to the quasar \PKS1830. We used ALMA to observe several submillimeter lines of ND, NH$_2$D, and HDO, and their related isotopomers NH$_2$, NH$_3$, and H$_2^{18}$O, in absorption against the southwest image of the quasar, allowing us to derive XD/XH abundance ratios. The absorption spectra mainly consist of two distinct narrow velocity components for which we find remarkable differences. One velocity component shows XD/XH abundances that is about 10 times larger than the primordial elemental D/H ratio, and no variability of the absorption profile during the time span of our observations. In contrast, the other component shows a stronger deuterium fractionation. Compared to the first component, this second component has XD/XH abundances that are 100 times larger than the primordial D/H ratio, a deepening of the absorption by a factor of two within a few months, and a rich chemical composition, with relative enhancements of N$_2$H$^+$, CH$_3$OH, SO$_2$ and complex organic molecules. We therefore speculate that this component is associated with the analog of a Galactic dark cloud, while the first component is likely more diffuse. Our search for the $^7$LiH (1--0) line was unsuccessful and we derive an upper limit $^7$LiH/H$_2$ $= 4 \times 10^{-13}$ (3$\sigma$) in the z=0.89 absorber toward \PKS1830. Besides, with ALMA archival data, we could not confirm the previous tentative detections of this line in the z=0.68 absorber toward B\,0218+357; we derive an upper limit $^7$LiH/H$_2$ $= 5 \times 10^{-11}$ (3$\sigma$), although this is less constraining than our limit toward \PKS1830. We conclude that, as in the Milky Way, only a tiny fraction of lithium nuclei are possibly bound in LiH in these absorbers at intermediate redshift.}

\maketitle

\section{Introduction}

Studying the evolution of elemental abundances and isotopic ratios in the Universe allows us to trace the history of nucleosynthesis. Predictions from the standard model of Big Bang nucleosynthesis (BBN; see, e.g., \citealt{cyb16}) set the theoretical abundances of light elements (i.e., the isotopes of H, He, and Li) given the baryon density, neutron lifetime, and some essential nuclear reaction rates. Heavier elements (C, N, O, ...) are subsequently produced by stellar nucleosynthesis.

There is no known astrophysical source of deuterium production so that the net evolution of the D/H ratio is expected to be a slow decrease by astration \citep{eps76}. Observations of high-redshift quasar absorption systems provide a measurement of the primordial D/H ratio of $(2.527 \pm 0.030) \times 10^{-5}$ (see \citealt{coo18} and reference therein), in agreement with nucleosynthesis models based on {\em Planck} cosmic microwave background (CMB) data \citep{PLANCK18}.

For lithium, measurements in the atmosphere of extreme metal-poor stars converge to a ``plateau'' value of Li/H = $(1.6 \pm 0.3) \times 10^{-10}$, which appears lower than predictions from BBN and CMB by a factor 3--4; this is known as the so-called lithium problem (e.g., \citealt{fie11}). A wide range of potential solutions to this problem have been proposed, for example through stellar physics (e.g., \citealt{bar17}) or with varying fundamental constants \citep{cla20}.

Measurements of D/H and Li/H abundances can also be done, in principle, using molecules, via abundance ratios of deuterated over hydrogenated isotopologues, XD/XH, and LiH/H$_2$, respectively. This, however, is complicated by astrochemical processes, which are interesting to study for their own sake. For example, the difference in molecular binding energy between the H and D isotopes of hydrogen can cause drastic enhancements of the isotopologue abundance ratios XD/XH over the isotopic ratio D/H, known as so-called chemical fractionation. Indeed, despite the low D/H ratio of $\sim 10^{-5}$, XD/XH abundance ratios of a few percent up to several tens of percent have been observed in various sources in the Milky Way and for different species, from the first detections of deuterated species (DCN; \citealt{jef73}, and HD, \citealt{spi73}) to the spectacular detections of triply-deuterated ammonia, ND$_3$ \citep{lis02,vdtak02}, and triply-deuterated methanol, CD$_3$OH \citep{par04}. As for extragalactic sources, deuterated molecules other than HD have been detected so far in the Large Magellanic Cloud (DCO$^+$, DCN; \citealt{chi96, hei97}) and, tentatively, in the starburst galaxy NGC253 (DNC, N$_2$D$^+$; \citealt{mar06}). Nevertheless, it is fair to say that as a consequence of the sensitivity of chemical fractionation to the temperature, studies of deuterated molecules usually serve more as diagnostics of the physical or chemical conditions, rather than to establish a D/H ratio.

In this work, we report the first detection of three deuterated species, ND, NH$_2$D, and HDO, and an upper limit to $^7$LiH, in the z=0.89 absorber against the quasar \PKS1830. We compare their absorption profiles and column densities to those of other species, in particular NH$_2$ and H$_2^{18}$O, and estimate the corresponding deuterium fractions, which we find up to $\sim 10^{-3}$ in a narrow and chemically peculiar velocity component.

\section{Observations} \label{sec:obs}

The data used in this study were observed with the Atacama Large Millimeter/submillimeter Array (ALMA) on 2019 April 11 and 2019 July 28, in the frame of two different science projects (2018.1.00692.S and 2018.1.00051.S). Table~\ref{tab:obsdata} gives a summary of the observations.

The correlator was configured to cover four spectral windows, each 1.875~GHz wide, with a resulting velocity resolution of 1--5~\kms\ (after Hanning smoothing), depending on the project. The data calibration was carried out within the Common Astronomy Software Applications package (CASA\footnote{http://casa.nrao.edu/}), following a standard procedure. The bandpass response of the antennas was calibrated from observations of the bright quasar J\,1924$-$292. At millimeter/submillimeter wavelengths, \PKS1830\ appears as two bright and compact continuum images (hereafter labeled as the NE and SW images) separated by $\sim 1\arcsec$. After the standard gain calibration using the quasar J\,1832$-$2039, the visibilities were further self-calibrated on the continuum of \PKS1830, by one iteration of phase-only solutions per 6 sec integration.

The final spectra were extracted using the CASA Python task UVMULTIFIT (\citealt{mar14}) by fitting a model of two point sources to the interferometric visibilities. The equivalent beam resolution of the data was largely sufficient to separate the two lensed images of \PKS1830, and this is even better done with visibility fitting, given the simple source geometry and the very high signal-to-noise ratio of the data. The observations on 2019 April 11 were done in full polarization mode (Marti-Vidal et al., in prep.) with two consecutive executions which were calibrated separately. The spectra, extracted from Stokes~I data, were then normalized to the continuum level and averaged.

We note that most of the z=0.89 absorption lines are seen only toward the SW image of \PKS1830\footnote{There is also some absorption toward the NE image, but only seen for the strongest absorption lines, and with a velocity shift of $\sim -150$~\kms, with respect to the peak of SW absorption (see, e.g., \citealt{mul14}).}, and in this work we only discuss this line of sight. Nevertheless, the spectra extracted from the NE image serve as a monitor to control the quality of the spectral bandpass, that is, to safely discard confusion with potential Galactic absorption, atmospheric lines, or bandpass ripples.

The list of lines mainly discussed in this paper is given in Table~\ref{tab:lines}. All velocities in this paper are referred to the heliocentric frame, taking a redshift of z=0.88582 \citep{wik96}.

\begin{table*}[ht!]
\caption{Summary of the ALMA observations of deuterated species toward \PKS1830.}
\label{tab:obsdata}
\begin{center} \begin{tabular}{lccccccc}
\hline \hline

Date & ALMA & N$_{\rm ant}$ & PWV & t$_{\rm on}$ & $\delta v$ & Beam  & Species \\
     & band &             & (mm) & (min)      & (\kms)     & (arcsec) & \\ 
\hline
2019 Apr 11 & B6 & 43 & 1.0 & 66 & 4.7 & 0.6 & NH$_2$D, HDO, p-NH$_2$, ($^7$LiH) \\ 
2019 Jul 28 & B7 & 42 & 0.6 &  19 & 1.1 &  0.03 & ND, o-H$^2_{18}$O \\ 
\hline
\end{tabular} \end{center}
\tablefoot{N$_{\rm ant}$: number of antennas in the array; PWV: precipitable amount of water vapor; t$_{\rm on}$: on-source integration time; $\delta v$: velocity resolution after Hanning smoothing.}
\end{table*}

\begin{table*}[ht]
  \caption{List of the observed lines of deuterated species and related species.}
\label{tab:lines}
\begin{center} \begin{tabular}{cllcc}
    \hline
    \hline
Line & Rest freq. & Redshifted freq.$^\triangle$   & $\mu$ $\ddagger$ & $\alpha$ $^\star$ \\
     & (GHz)      & (GHz) & (D) & (cm$^{-2}$\,km$^{-1}$\,s) \\
\hline
\multicolumn{5}{c}{\em Observed on 2019 April 11} \\
\hline
HDO ($J_{Ka},_{Kc}$=$1_{0,1}$--$0_{0,0}$)  & 464.924520 & 246.537  & 1.73 & $1.9 \times 10^{13}$ \\  
o-NH$_2$D ($J_{Ka},_{Kc}$=$1_{1,0}$--$0_{0,0}$, $F$=2--1) & 470.271911$^\dagger$ & 249.373  & 1.46 & $5.6 \times 10^{13}$$^\diamondsuit$$^\circ$ \\
p-NH$_2$ ($N_{Ka},_{Kc}$=$1_{1,0}$--$1_{0,1}$, $J$=\nicefrac{1}{2}--\nicefrac{1}{2}, $F$=\nicefrac{3}{2}--\nicefrac{3}{2}) & 469.440621$^\dagger$ & 248.932$^\dagger$ & 1.82 & $1.8 \times 10^{14}$$^\diamondsuit$$^\circ$ \\
p-NH$_2$ ($N_{Ka},_{Kc}$=$1_{1,0}$--$1_{0,1}$, $J$=\nicefrac{1}{2}--\nicefrac{3}{2}, $F$=\nicefrac{3}{2}--\nicefrac{5}{2}) & 470.409052$^\dagger$ & 249.445$^\dagger$ & 1.82 & $1.8 \times 10^{14}$$^\diamondsuit$$^\circ$ \\
$^7$LiH ($J$=1--0) & 443.952930 &  235.416 & 5.88 & $2.5\times 10^{11}$ \\

\hline
\multicolumn{5}{c}{\em Observed on 2019 July 28} \\
\hline
ND ($N$=1--0, $J$=2--1)   & 522.077372$^\dagger$ & 276.844$^\dagger$ & 1.39 & $7.7 \times 10^{13}$$^\diamondsuit$ \\
ND ($N$=1--0, $J$=1--1)   & 546.127619$^\dagger$ & 289.597$^\dagger$ & 1.39 & $7.7 \times 10^{13}$$^\diamondsuit$ \\
o-H$_2^{18}$O ($J_{Ka},_{Kc}$=$1_{1,0}$--$1_{0,1}$) & 547.676440 &  290.418 & 1.85 & $6.3 \times 10^{12}$$^\circ$ \\  

\hline
\end{tabular}\end{center}
\tablefoot{$\triangle$ For a redshift z=0.88582, in heliocentric frame; $\ddagger$ electric dipole moment; $\star$ coefficients to convert line integrated opacity into column density $N_{\rm col} = \alpha \int \tau dv$, calculated in the case of pure radiative coupling with CMB photons, i.e., taking $\Tex=\Tcmb=5.14$~K; $\dagger$ frequency of the strongest hyperfine transition; 
  $\diamondsuit$ for a hyperfine component with equivalent line strength $S_{ul}=1$;
  $\circ$ the coefficient converts directly to the total (ortho+para) column density, assuming the OPR=3. Spectroscopy parameters are taken from the Cologne Database for Molecular Spectroscopy \citep{CDMS01,CDMS05,CDMS16} and the Jet Propulsion Laboratory Molecular Spectroscopy catalog \citep{pic98} and references therein.}
\end{table*}

\begin{figure}[h] \begin{center}
\includegraphics[width=8.8cm]{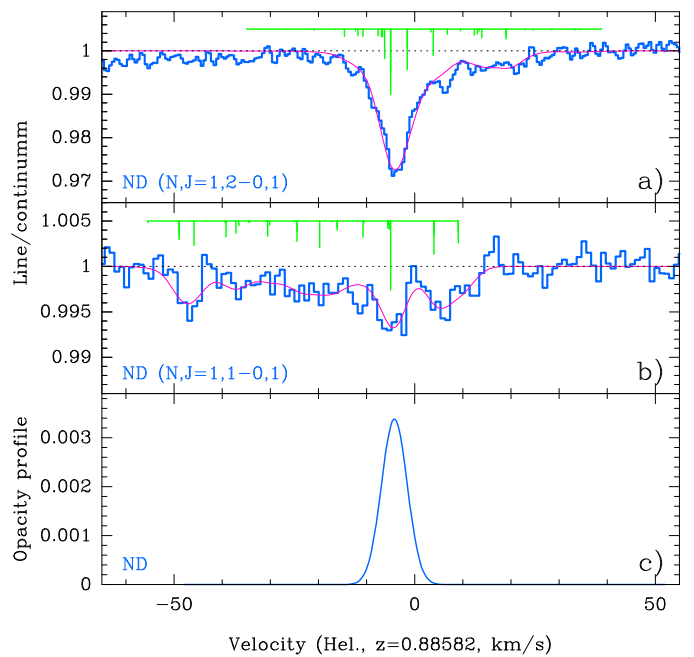}
\caption{Spectra of the ND ($N$=1--0, $J$=2--1) ($a$) and ($N$=1--0, $J$=1--1) ($b$) lines toward the SW image of \PKS1830, observed with ALMA in July 2019. The hyperfine structure and the best fits with one single Gaussian velocity component are shown. c) Corresponding ND opacity profile fitted with one Gaussian velocity component (see Table~\ref{tab:Gaussfit}), normalized for an equivalent transition of line strength =1.}
\label{fig:spec-ND}
\end{center} \end{figure}

\begin{figure}[h] \begin{center}
\includegraphics[width=8.8cm]{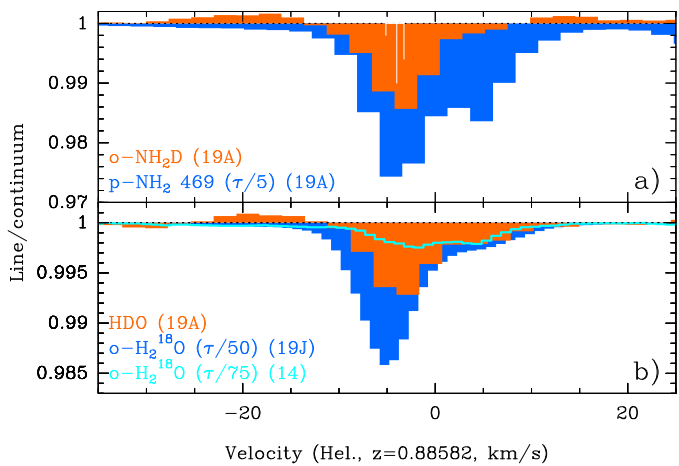}
\caption{Spectra of the ortho-NH$_2$D ($1_{1,0}$--$0_{0,0}$) and para-NH$_2$ ($1_{1,0}$--$1_{0,1}$ J=$\nicefrac{1}{2}$--$\nicefrac{1}{2}$, F=$\nicefrac{3}{2}$--$\nicefrac{3}{2}$) ($a$) and HDO ($1_{0,1}$--$0_{0,0}$) and ortho-H$_2^{18}$O ($1_{1,0}$--$1_{0,1}$) ($b$) lines observed with ALMA toward the SW image of \PKS1830. The labels in parenthesis (14), (19A), and (19J) stand for observations done in 2014, 2019 April, and 2019 July, respectively. The spectra of p-NH$_2$, o-H$_2^{18}$O(19J), and o-H$_2^{18}$O(14) were scaled down in opacity by factors 5, 50, and 75, respectively. The hyperfine structure of the ortho-NH$_2$D line (see Table\,\ref{tab:spectro-NH2D}) is indicated, with the strongest hfs component centered at a velocity of $-4$~\kms.}
\label{fig:spec-deuterated}
\end{center} \end{figure}

\section{Results and analysis}

\subsection{Analysis of deuterated species}

With these ALMA data, we detect three different deuterated species: ND (two groups of fundamental hyperfine structure transitions, $N$=1--0, $J$=2--1 at 522~GHz and $N$=1--0, $J$=1--1 at 546~GHz, rest frequencies), ortho-NH$_2$D ($1_{1,0}$--$0_{0,0}$ fundamental transition at 470 GHz), and HDO ($1_{0,1}$--$0_{0,0}$ fundamental transition at 464.9 GHz). Their absorption spectra are shown in Figs. \ref{fig:spec-ND} and \ref{fig:spec-deuterated}.

\begin{figure}[h] \begin{center}
\includegraphics[width=8.8cm]{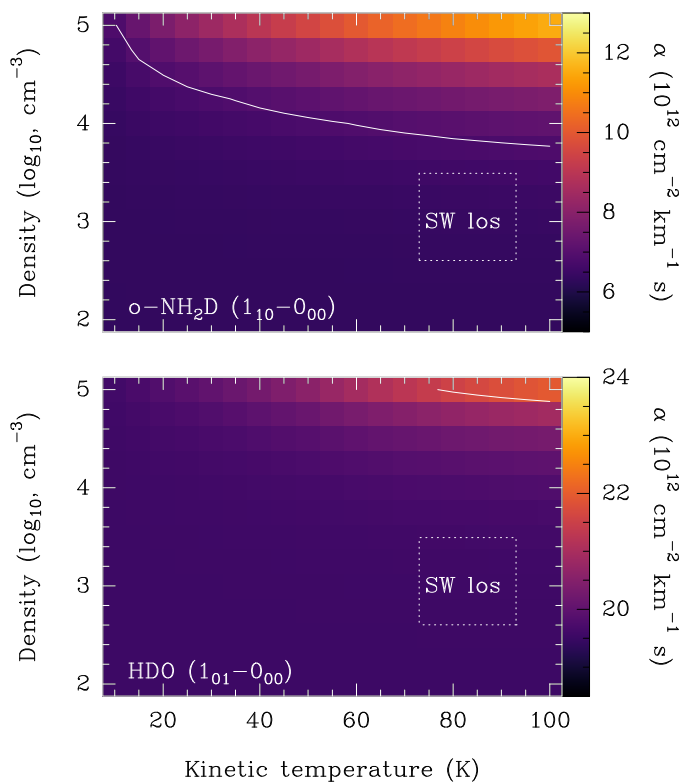}
\caption{Conversion factors $\alpha$ between integrated opacity to column density derived with RADEX for the fundamental transitions of NH$_2$D and HDO, for a range of standard conditions in the interstellar medium. We assume the OPR=3. For the o-NH$_2$D ($1_{1,0}$--$0_{0,0}$) transition, the conversion factor corresponds to the calculation with a collapsed hyperfine structure, i.e., 9 times lower than for an equivalent component with line strength of 1 (Tables~\ref{tab:lines} and \ref{tab:spectro-NH2D}). The white curve indicates a departure of 10\% over the pure radiative equilibrium calculation for $T_{\rm ex}=\Tcmb=5.14$~K. The white dotted box labeled ``SW los'' indicates the range of densities and temperatures previously estimated in the line of sight to the SW image of \PKS1830\ by \cite{mul13}.}
\label{fig:alpha}
\end{center} \end{figure}

In order to convert the observed integrated opacities into column densities, we need to make some assumptions. Since all species have large electric dipole moments, $> 1$~D, we expect that their excitation is locked to the temperature of the CMB (see \citealt{mul13}). We define the conversion factors $\alpha$ as $N_{\rm col} = \alpha \int \tau dv$. In the case of pure radiative equilibrium with CMB photons, the $\alpha$ factors can be calculated with $\Tcmb =5.14$~K at z=0.89 (again, see \citealt{mul13}), and their values are given in Table~\ref{tab:lines}. In addition, we ran the radiative transfer code RADEX \citep{vdtak07} for NH$_2$D and HDO, for which preformatted collisional rates are available in the LAMDA database \citep{sch05}, to check the effects of collisional excitation in standard physical conditions in the interstellar medium. The rates are taken from \cite{dan14} for NH$_2$D and from \cite{fau12} for HDO. We assume only H$_2$ as collisional partner. We see in Fig.\,\ref{fig:alpha} that indeed the corresponding $\alpha$ conversion factors remain constant within 10\% for volume density up to $10^4$~cm$^{-3}$ and kinetic temperature in the range 10--100~K. 

We further assume that the absorbing gas completely and uniformly covers the background continuum emission from the quasar, that is, a covering factor $\fc=1$. This is supported by the fact that heavily saturated lines of CO, HCO$^+$, HCN, and H$_2$O have already been shown to reach absorption depths close to 100\% toward the SW image of \PKS1830\ \citep{mul14}. Concerning the present dataset, other lines observed within the same tunings as the deuterated species help to clarify this point. The depth of the o-H$_2^{18}$O ($1_{1,0}$--$1_{0,1}$) and N$_2$H$^+$ (3-2) lines observed in July 2019 reaches a level of $\sim 50$\% of the continuum (Fig.\,\ref{fig:spec-timevar}), that is, de facto, $\fc > 50$\%. Besides, the absorption profile of p-NH$_2$ reaches a depth of $\sim 15$\% of the continuum level and follows well the relative intensities of the hyperfine structure (hfs) expected when the sublevels are populated in proportion to their statistical weights (Fig.\,\ref{fig:spec-NH2}). These facts suggest that the p-NH$_2$ lines are indeed optically thin, and we therefore expect that the assumption $\fc=1$ has a negligible impact on the measurements of column densities of lines weaker than that. 

Since we are mainly interested in the XD/XH abundance ratios (i.e., deuterated versus main isotopologues), we would need to compare ND, NH$_2$D, and HDO to the corresponding NH, NH$_3$, and H$_2$O, respectively. However, all the D-species were discovered as secondary products of two projects with different science goals, not specifically targeting their main isotopologues. The first fundamental hyperfine transitions of NH ($\sim 946$ and 974~GHz) could in principle be observed from the ground under good weather conditions, but are redshifted just beyond the edge of ALMA band~8 for z=0.89. Therefore NH is out of reach in \PKS1830's absorber for the moment. As for both NH$_3$ and H$_2$O, the fundamental transition of their ortho form has been observed previously with ALMA in 2012 (e.g., \citealt{mul14,mul16b}). However, known time variations of the absorption profiles due to structural changes in the morphology of the background quasar (\citealt{mul08,mul14,sch15}) preclude us from using these past observations directly without careful comparison of reference spectra. In addition, the o-H$_2$O transition is heavily saturated, and we need to turn to a rare isotopologue to get a proper line opacity measurement, modulo the value of the isotopic ratio. Fortunately, several other molecules, including NH$_2$ and H$_2^{18}$O, have transitions detected simultaneously within the frequency range of our observations. Hereafter, we use these complementary spectra to unlock the XD/XH abundance ratios.

\subsubsection{NH$_2$ as a reference to unlock the abundances of nitrogen hydrides in our data}

\begin{figure}[h] \begin{center}
\includegraphics[width=8.8cm]{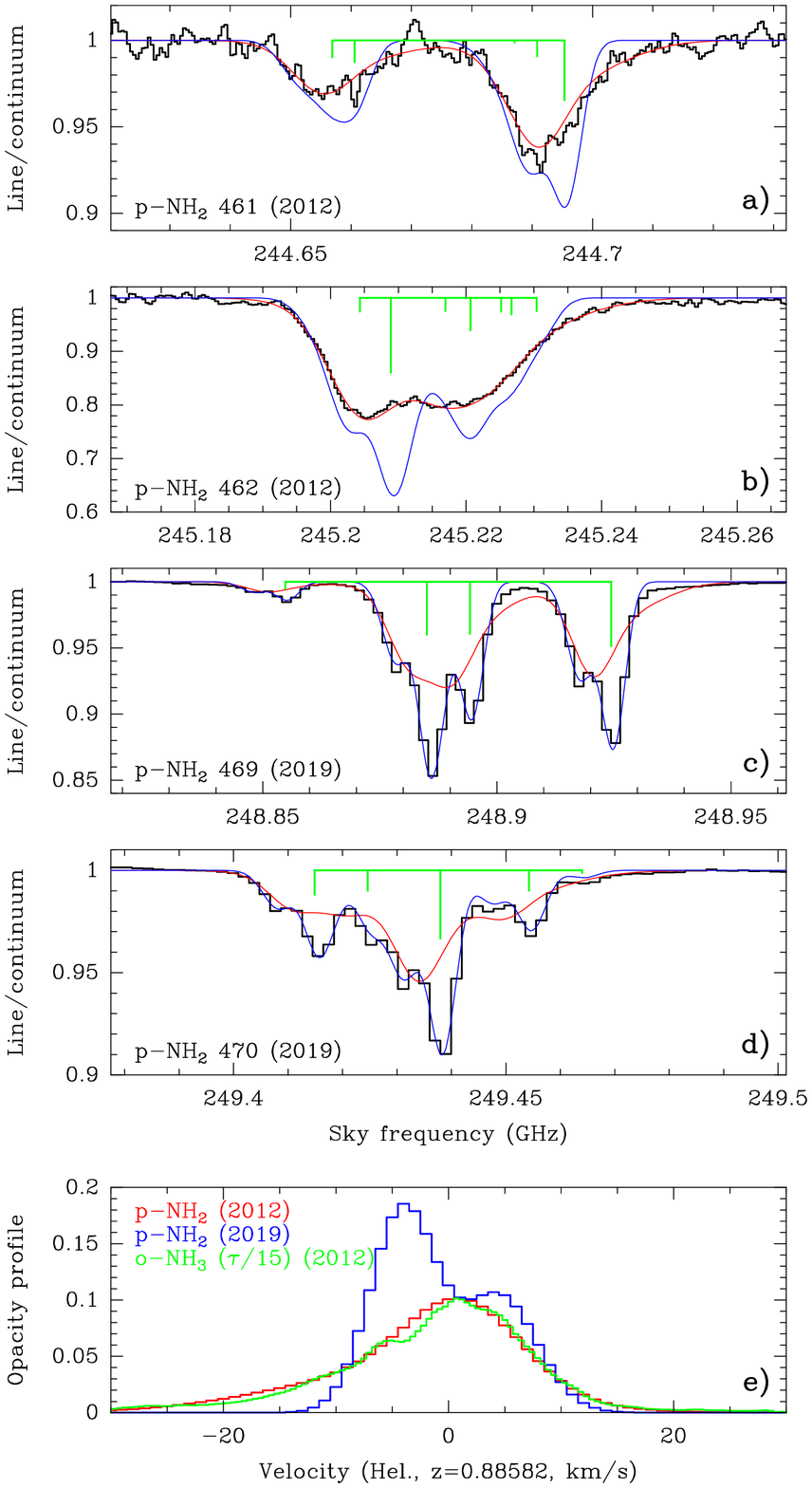}
\caption{$a,b,c,d$) Spectra of the different groups of fundamental hyperfine structure transitions of the para-NH$_2$ ($1_{1,0}$--$1_{0,1}$) line observed with ALMA at different epochs toward the SW image of \PKS1830. The hyperfine structure is shown on top of each group in green. The 2012 and 2019 profiles were fitted each separately with two Gaussian velocity components and the synthetic spectra corresponding to the best fits are shown in red (matching the observed spectrum in 2012) and blue (matching the observed spectrum in 2019). $e$) The last panel shows the 2012 and 2019 opacity profiles in velocity fitted for p-NH$_2$ (i.e., profiles deconvolved from the hyperfine structure and normalized for an equivalent transition of line strength =1), with the same color code as in $a$--$d$), as well as the profile of the o-NH$_3$ ($1_0$--$0_0$) transition (in green) observed in 2012 by \cite{mul14}, scaled to the peak of the p-NH$_2$ 2012 profile.}
\label{fig:spec-NH2}
\end{center} \end{figure}

Two groups of fundamental hyperfine structure transitions of para-NH$_2$ were observed simultaneously with NH$_2$D on 2019 April 11, and two others were observed together with ortho-NH$_3$ back in 2012 \citep{mul14}; these detections made it possible to determine the ND/NH and NH$_2$D/NH$_3$ ratios. These spectra are compiled in Fig.\,\ref{fig:spec-NH2}. The spectroscopy of NH$_2$ is complicated by the geometry of the molecule, unpaired electron spin, ortho/para states due to two hydrogen nuclei, and hyperfine splitting due to the nitrogen atom (see, e.g., \citealt{vandis93}). The present spectra concern para-NH$_2$, so that the nuclear spin hfs reduces to that of nitrogen. A study of NH$_2$ excitation was recently presented by \cite{bou19}. The critical density of the transition observed by us is on the order of $10^8$--$10^9$~cm$^{-3}$ so that in our case the excitation temperature remains close to the CMB temperature. Hence, as for HDO and NH$_2$D, we assume that the $\alpha$ opacity-to-column density conversion factors for NH$_2$ transitions are equal to the values calculated for $\Tex = \Tcmb$ (Table\,\ref{tab:lines}).

We can deconvolve the 2012 and 2019 absorption profiles in a satisfactory way with two Gaussian velocity components each (Fig.\,\ref{fig:spec-NH2}e). We observe clear time variations between the 2012 and 2019 profiles, with, especially, an increase by a factor $\sim 2$ of the integrated opacity between $v=-10$~\kms\ and $v=0$~\kms\ in 2019. On the other hand, the absorption profiles of p-NH$_2$ and o-NH$_3$ showed a good match in 2012 (Fig.\,\ref{fig:spec-NH2}e) and yield a total column density $\sim 3 \times 10^{14}$~cm$^{-2}$ for both, that is, NH$_2$/NH$_3$ $\sim 1$, after correction for the statistical ortho-to-para ratios (OPRs) of three for NH$_2$ and one for NH$_3$. Although OPRs can deviate from their statistical values because they depend on the formation temperature of the molecules and can be modified by chemical reactions, we note that the observed deviations are moderate, where OPR=2--4 for NH$_2$ and 0.5--1 for NH$_3$ (e.g., \citealt{per16,legal16}).

Our NH$_2$/NH$_3$ ratio resembles the abundance sequence for nitrogen hydrides NH:NH$_2$:NH$_3$= 3--2:1:1 observed in the Galactic diffuse gas \citep{per10,per12}, and is in stark contrast with that found in dark dense clouds, 3:1:19, where the abundance of ammonia seems to be enhanced by about one order of magnitude with respect to NH and NH$_2$ \citep{legal14}. This suggests a diffuse nature for the absorbing gas in the z=0.89 absorber. We emphasize, however, that our measurement NH$_2$/NH$_3$ $\sim 1$ is derived for the 2012 line of sight and does not necessarily hold in 2019, since it is not impossible that a different gas component had been illuminated at that time.

A simple comparison of the absorption profiles of p-NH$_2$ and o-NH$_2$D (Fig.\,\ref{fig:spec-deuterated}a) reveals that they are not similar. The absorption of o-NH$_2$D is clearly detected for the $v\sim-4$~\kms\ velocity component, but falls in the noise at $v\sim +5$~\kms, where p-NH$_2$ still shows strong absorption. We fit the different spectra with Gaussian velocity components and the best-fit results are given in Table~\ref{tab:Gaussfit}. Since o-NH$_2$D and p-NH$_2$ were observed simultaneously, we can directly extract the abundance ratio NH$_2$D/NH$_2$ = $(2.7 \pm 0.3)\times 10^{-3}$ for the $v\sim-4$~\kms\ velocity component. Further assuming NH$_2$/NH$_3$=1, as we have discussed above, we end up with the ratio NH$_2$D/NH$_3$ $\sim 3 \times 10^{-3}$. This corresponds to a deuterium fractionation of $\sim 100\times$ over the primordial D/H ratio. Taking an NH$_2$/NH$_3$ ratio of 19, as observed for dark dense clouds \citep{legal14}, would still yield a significant deuterium fractionation, $\sim 10\times$ primordial. On the other hand, with the fit results for the $v=+5$~\kms\ component (Table~\ref{tab:Gaussfit}) and NH$_2$/NH$_3$=1, we find a ratio NH$_2$D/NH$_3$ $\sim 4 \times 10^{-4}$, lower by almost one order of magnitude compared to the $v=-4$~\kms\ component.

Now turning to ND for the $v=-4$~\kms\ component, we see that its one Gaussian velocity component fit is consistent in velocity centroid and line width compared to the values found for NH$_2$D. First assuming no time variations between April and July 2019, we derive ND/NH$_2$D = $2.7 \pm 0.2$, in good agreement with the Galactic diffuse NH/NH$_3$ ratio, which would imply a similar degree of deuterium fractionation for NH and NH$_3$. Even if the column density of ND might have increased by a factor $\sim 2$ between April and July 2019, as we discuss in the following section, we still get an estimate of ND/NH$_2$D near unity. With the assumption NH/NH$_2$ = 3, we further derive ND/NH $= (2.44 \pm 0.07) \times 10^{-3}$ for the $v\sim-4$~\kms\ velocity component\footnote{The errors come from the fit uncertainties only, and include neither potential time variations nor the uncertainty on the NH/NH$_2$ ratio.}, consistent with the deuterium fractionation obtained from NH$_2$D/NH$_3$.

To get an estimate of ND/NH for the $v=+5$~\kms\ component, we forced the fit of the ND spectrum with the same two Gaussian velocity components obtained from the fit of p-NH$_2$ (Table~\ref{tab:Gaussfit}). Assuming no time variations, and NH/NH$_2$ = 3, we obtain a ratio ND/NH = $(1.3 \pm 0.8) \times 10^{-4}$, again roughly one order of magnitude lower than for the $v=-4$~\kms\ component.

\begin{table*}[ht!]
\caption{Results of Gaussian fitting.}
\label{tab:Gaussfit}
\begin{center} \begin{tabular}{lcccccccc}
\hline \hline
        &      & \multicolumn{3}{c}{Velocity component 1} & \multicolumn{3}{c}{Velocity component 2} & Notes \\
Species & Date & $v_1$ & FWHM$_1$ & $\int \tau dv$ & $v_2$ & FWHM$_2$ & $\int \tau dv$ & \\
        &      & (\kms) & (\kms) & (\kms) & (\kms) & (\kms) & (\kms) & \\
\hline
p-NH2   & 2012 & $-6.0 \pm 0.8$ & $24.3 \pm 0.8$ & $0.81 \pm 0.08$ & $1.3 \pm 0.1$ & $12.2 \pm 0.4$ & $0.99 \pm 0.08$ & 2G \\
\hline
p-NH2   & 2019/04/11 & $-4.02 \pm 0.06$ & $6.9 \pm 0.1$ & $1.34 \pm 0.03$ $^\dagger$ & $4.5 \pm 0.1$ & $7.9 \pm 0.2$ & $0.87 \pm 0.03$ $^\dagger$ & 2Gt \\
o-NH$_2$D & & -- & -- & $(11.5 \pm 1.2)$$\times$$10^{-3}$ $^\dagger$ & -- & -- & $(1.0 \pm 1.3)$$\times$$10^{-3}$ $^\dagger$ & 2Gt \\
HDO & & -- & -- & $(52 \pm 11)$$\times$$10^{-3}$ & -- & -- & $(17 \pm 12)$$\times$$10^{-3}$ & 2Gt \\
\hline
o-NH$_2$D & 2019/04/11 & $-3.5 \pm 0.2$ & $6.8 \pm 0.4$ & $(11.7 \pm 0.6)$$\times$$10^{-3}$ $^\dagger$ & x & x & x & 1G\\
\hline
HDO     & 2019/04/11 & $-3.8 \pm 0.4$ & $6.4 \pm 0.8$ & $(50 \pm 8)$$\times$$10^{-3}$ & $4.8 \pm 1.9$ & $8.7 +/- 4.0$ & $(18 \pm 8)$$\times$$10^{-3}$ & 2G\\
\hline
ND      & 2019/07/28 & $-4.2 \pm 0.1$ & $6.3 \pm 0.2$ & $(22.7 \pm 0.4)$$\times$$10^{-3}$ $^\dagger$ & x & x & x & 1G \\
ND & 2019/07/28 & $-4.0$ $^\diamond$ & 6.9 $^\diamond$ & $(22.7 \pm 0.5)$$\times$$10^{-3}$ $^\dagger$ & 4.5 $^\diamond$ & 7.9 $^\diamond$ & $(0.8 \pm 0.5)$$\times$$10^{-3}$ & 2G \\
\hline
o-H$_2^{18}$O & 2019/07/28 & $-5.15 \pm 0.02$ & $5.71 \pm 0.04$ & $3.42 \pm 0.03$ & $-1.1 \pm 0.1$ & $17.6 \pm 0.2$ & $3.22 \pm 0.04$ & 2G\\
\hline
\end{tabular} \end{center}
\tablefoot{1G,2G: fit with one or two Gaussian velocity component(s), respectively; 2Gt: all species fitted simultaneously with the same tight 2-Gaussian velocity profile; $\dagger$ The integrated opacity is given for an equivalent hyperfine component of line strength $S_{ul}=1$. To obtain column density, multiply the integrated opacity by the corresponding conversion factor $\alpha$ in Table~\ref{tab:lines}. $\diamond$ Fixed parameter.}
\end{table*}

\subsubsection{HDO}

It is, unfortunately, more difficult to obtain a reliable D/H ratio from HDO. H$_2^{18}$O was observed in July 2019, that is, several months after HDO. The absorption profiles of the two species (Fig.\,\ref{fig:spec-deuterated}b) appear to be slightly different; the peak of the $v=-4$~\kms\ velocity component is apparently slightly shifted, beyond what we could expect owing to the differences in velocity resolution and signal-to-noise ratio between the two spectra only. The two profiles cannot be reproduced in a satisfactory way by a joint fit.

\begin{figure}[h] \begin{center}
\includegraphics[width=8.8cm]{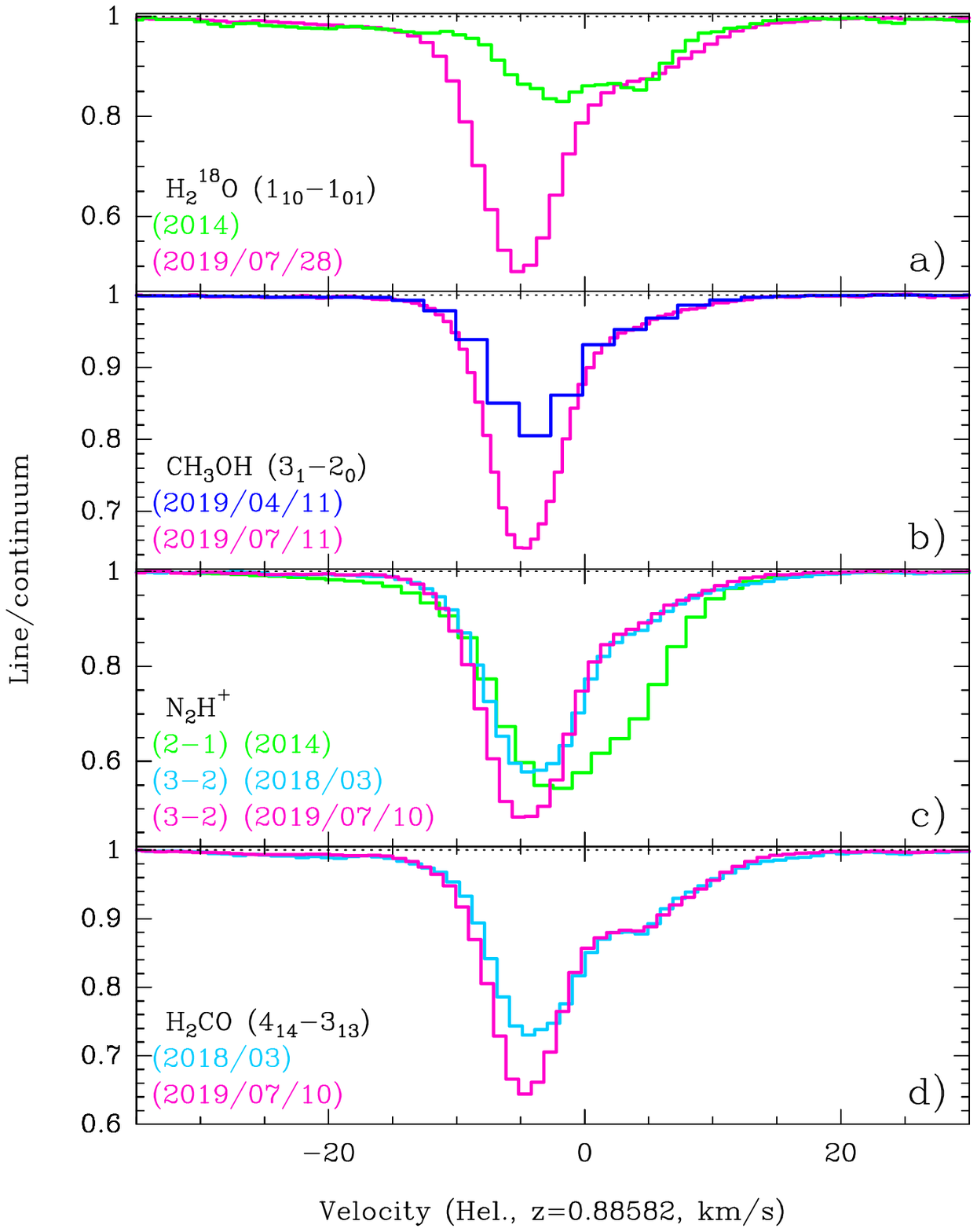}
\caption{Spectra of the H$_2^{18}$O ($1_{1,0}$--$1_{0,1}$), CH$_3$OH ($3_1$--$2_0$), N$_2$H$^+$ (2--1) and (3--2), and H$_2$CO ($4_{14}$--$3_{13}$) lines observed at different epochs with ALMA toward the SW image of \PKS1830, illustrating the time variations of the absorption profile.}
\label{fig:spec-timevar}
\end{center} \end{figure}

Several observations point toward a significant variability of the $v=-4$~\kms\ velocity component. We saw the case of p-NH$_2$ in the previous section. The fundamental transition of o-H$_2^{18}$O has already been observed in 2014 \citep{mul16b}, and we compare the two spectra in Fig.\,\ref{fig:spec-timevar}a. At that time, the profile was mostly reproduced by two velocity components, at $v \sim -5$~\kms\ and $v \sim +5$~\kms, 
of roughly equal optical depth. While the part of the profile with $v>0$~\kms\ barely changed, the line integrated opacity between $v=-10$~\kms\ and $v=0$~\kms\ increased by a factor of four between 2014 and 2019. Furthermore, the $3_1$--$2_0$ transition of methanol at a rest frequency of 445.571~GHz was also observed by ALMA in 2019 April 11 and 2019 July 11 (Fig.\,\ref{fig:spec-timevar}b). The line integrated opacity between $v=-10$~\kms\ and $v=0$~\kms\ increased by a factor two within a few months. The time variations of the absorption profile are further illustrated in Fig.\,\ref{fig:spec-timevar}c, where we collect spectra of N$_2$H$^+$ lines obtained in 2014, 2018, and 2019. Therefore, we can conclude that the $v \sim -5$~\kms\ component has been affected by significant variations between 2014 and 2019 with an apparent global increase of its optical depth. The observations are scarce, but assuming that the lines of H$_2^{18}$O and CH$_3$OH follow each other, we estimate a variation by a factor two of the integrated opacity between April and July 2019.

A second complication to derive HDO/H$_2$O from the HDO/H$_2^{18}$O abundance ratio arises with the required correction from the $^{16}$O/$^{18}$O isotopic ratio. Fractionation effects are expected to be negligible for H$_2^{18}$O (e.g., \citealt{loi19}), and we adopt the ratio $^{16}$O/$^{18}$O $= 69 \pm 11$ derived from observations of HCO$^+$ isotopologues by \cite{mul06,mul11}. We recall that the $^{16}$O/$^{18}$O ratio measured in the z=0.89 absorber is lower that the value in the local interstellar medium and solar system (e.g., \citealt{luc98, lod03}) by nearly one order of magnitude, but comparable to the ratio measured in another molecular absorber at z=0.68 \citep{wal16}.

Taking the integrated opacities for the $v=-4$~\kms\ component from Table~\ref{tab:Gaussfit} and applying the time variation and isotopic ratio corrections discussed above, we end up with a ratio HDO/H$_2$O $\sim 1 \times 10^{-3}$ that is comparable to the deuterium fractionation measured from the nitrogen species. Now for the $v\sim+5$~\kms\ component, we see from Fig.\,\ref{fig:spec-deuterated}b that the opacity ratio of the HDO to o-H$_2^{18}$O profiles is $\sim 75$. Assuming that this velocity component did not vary significantly between 2014 and 2019, as suggested by the spectra shown in Fig.\,\ref{fig:spec-timevar}, we estimate a ratio HDO/H$_2$O $\sim 6 \times 10^{-4}$. Although the uncertainties are high, the two velocity components seem to have slightly different deuterium fractionations.

\begin{table*}[ht!]
\caption{Summary of the estimated XD/XH ratios toward the $v=-4$~\kms\ component, and their associated uncertainties.}
\label{tab:ratio}
\begin{center} \begin{tabular}{llllll}
\hline \hline
Ratio A/B & Observed ratio & Pivot species & OPR & Time variations $^{\dagger}$ & Final ratio A/B \\
\hline
ND/NH          & ND/p-NH$_2$ & NH/NH$_2$=1--5 $^{\circ}$ & OPR(NH$_2$)=3 & $\times 2$ & (0.7--7)$\times 10^{-3}$\\
NH$_2$D/NH$_3$ & o-NH$_2$D/p-NH$_2$ & NH$_2$/NH$_3$=1--20 $^{\circ}$ & OPR(NH$_2$)=3 & -- & (1.4--27)$\times 10^{-4}$\\
HDO/H$_2$O     & HDO/o-H$_2^{18}$O & $^{16}$O/$^{18}$O=$69 \pm 11$ $^{\diamond}$ & OPR(H$_2$O)=3 & $\times 2$ & $\sim 1 \times 10^{-3}$ \\
\hline
ND/NH$_2$D & ND/o-NH$_2$D & -- & OPR(NH$_2$D)=3 & $\times 2$ & 1--3\\
o-NH$_2$D/p-NH$_2$ & -- & -- & OPR(NH$_2$D)=3 & -- & $2.7 \times 10^{-3}$ \\
                   &    &   & OPR(NH$_2$)=3  &     & \\
\hline
\end{tabular} \end{center}
\tablefoot{ $\dagger$ Time variations of the $v=-4$~\kms\ component between 2019 April 11 and 2019 July 11, inferred from observations of the same transition of CH$_3$OH, see spectra in Fig.\,\ref{fig:spec-timevar}; $\circ$ see \cite{per10,per12,legal14}; $\diamond$ see \cite{mul06,mul11}.}
\end{table*}

\subsection{Comparison with profiles of other species}

\begin{figure}[h] \begin{center}
\includegraphics[width=8.8cm]{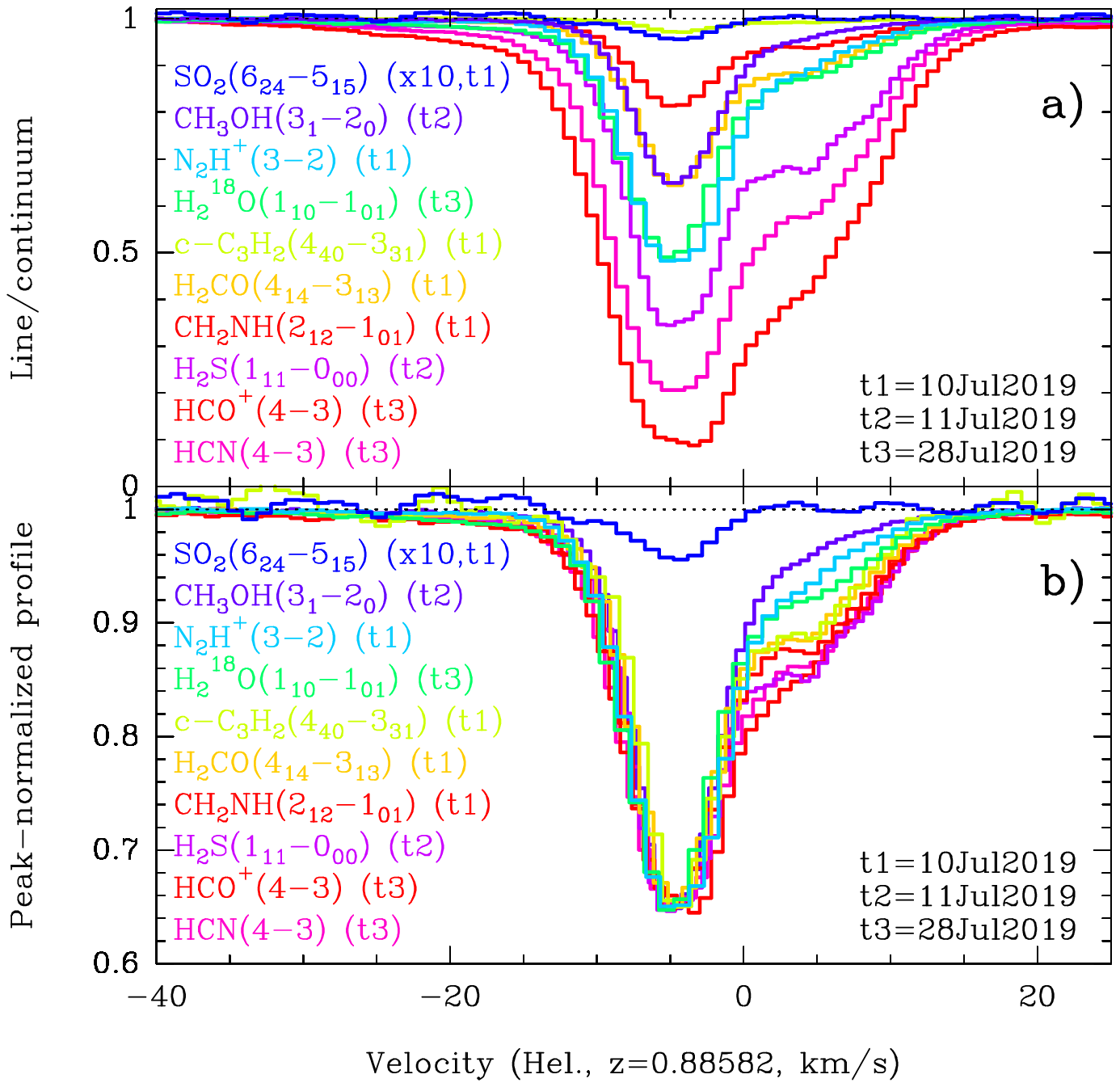}
\caption{a) Spectra of species observed with ALMA in July 2019 toward the SW image of \PKS1830. b) Same as before but all profiles have been normalized to the peak of the CH$_3$OH absorption, except that of the SO$_2$($6_{24}$--$5_{15}$) line, which was multiplied by a factor 10 in opacity.}
\label{fig:spec-comparo-profiles}
\end{center} \end{figure}

Chemical differentiation between velocity components was already noticed by \cite{mul16a}, with CF$^+$ and C absorption peaking at $v \sim +5$~\kms, in contrast to $^{13}$CO and CH$_3$OH peaking at $v=-5$~\kms. The spectra of HCO$^+$, HCN, H$_2$S, H$_2$CO, c-C$_3$H$_2$, CH$_2$NH, H$_2^{18}$O, N$_2$H$^+$, and CH$_3$OH were observed concomitantly with the D-species in July 2019, providing us with interesting clues (Fig.\,\ref{fig:spec-comparo-profiles}). We also add the detection of the SO$_2$($6_{24}$--$5_{15}$) line, which represents the first detection of SO$_2$ toward \PKS1830. For comparison of the species, we normalized their profiles to the $v=-4$~\kms\ peak \footnote{We note that we assumed the same constant covering factor over the entire profile.}. We can clearly define two groups based on the absorption depth ratio of the $v=-5$~\kms\ to $v \sim +5$~\kms\ component. Similar to the D-species and with a clear enhancement of their absorption toward $v \sim -5$~\kms\ and weak absorption at velocity $v \sim +5$~\kms, we find CH$_3$OH, N$_2$H$^+$, H$_2^{18}$O and  SO$_2$, although the latter has limited signal-to-noise ratio. On the other hand, HCO$^+$, HCN, and H$_2$S, which are known to be more ubiquitous in the diffuse gas component (e.g., \citealt{luc96}) have a deeper absorption component at velocity $v \sim +5$~\kms\ relative to the former group. CH$_2$NH, H$_2$CO, and c-C$_3$H$_2$ are in an intermediate situation.

\subsection{Non-detection of $^7$LiH}

\begin{figure}[h] \begin{center}
\includegraphics[width=8.8cm]{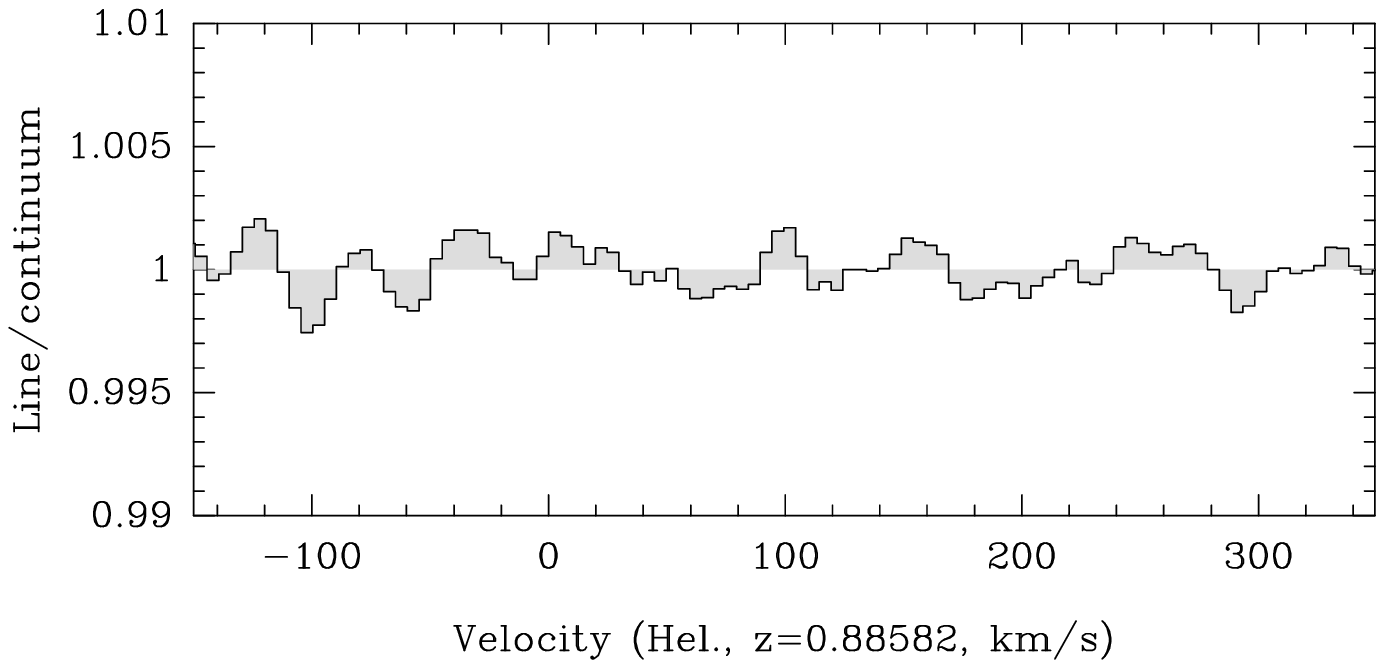}
\caption{Non-detection of $^7$LiH (1--0) toward the SW image of \PKS1830.}
\label{fig:spec-7LiH-PKS}
\end{center} \end{figure}

The spectrum of the $^7$LiH (1--0) line is shown in Fig.\,\ref{fig:spec-7LiH-PKS} and represents a non-detection. An upper limit on the integrated opacity of $^7$LiH can be calculated as $3\sigma_{\tau} \sqrt{\delta v \Delta v}$, where $\sigma_{\tau} = 1 \times 10^{-3}$ is the root-mean-square\ (rms) noise level normalized to the continuum level per velocity channel $\delta v = 4.7$~\kms, and $\Delta v$ is the line width. To be conservative we take $\Delta v =20$~\kms. With a conversion factor of $\alpha=2.5 \times 10^{11}$~cm$^{-2}$\,km$^{-1}$\,s, we then obtain an upper limit of $7 \times 10^9$~cm$^{-2}$ for the column density of $^7$LiH. Taking the column density of H$_2$ estimated along the SW line of sight~\footnote{This column density of H$_2$ might also be subject to a factor of 2--4 uncertainty owing to the time variability of the source.}, $2 \times 10^{22}$~cm$^{-2}$ \citep{mul14}, we derive a stringent upper limit $^7$LiH/H$_2$ $< 4 \times 10^{-13}$ (3$\sigma$).

\begin{figure}[h] \begin{center}
\includegraphics[width=8.8cm]{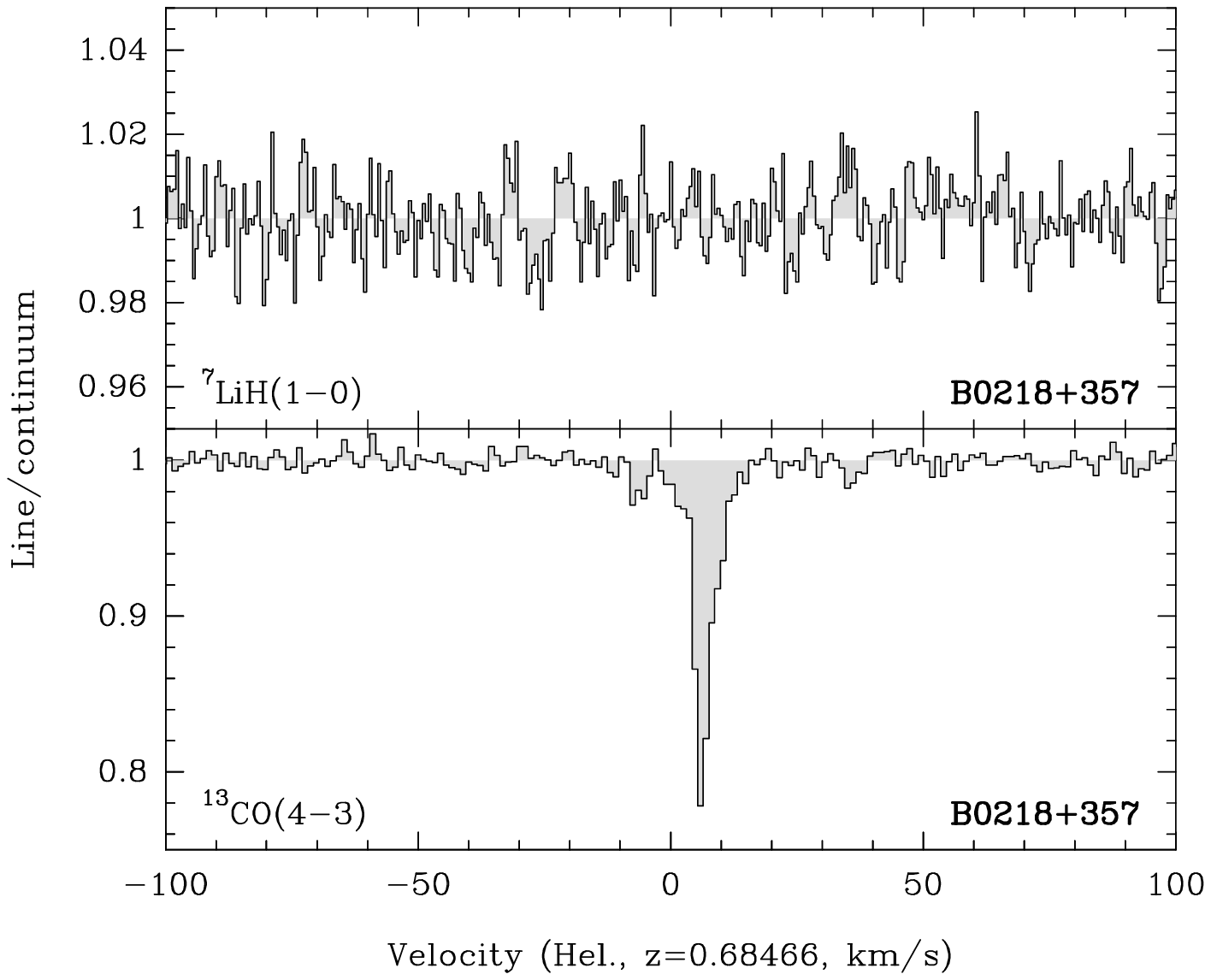}
\caption{Spectra of $^7$LiH (1--0) and $^{13}$CO (4--3) observed toward B\,0218+357 with ALMA.}
\label{fig:spec-7LiH-B0218}
\end{center} \end{figure}

\cite{com98} and \cite{fri11} both reported independent tentative detections of $^7$LiH (1--0) absorption in the z=0.68 absorber toward B\,0218+357 with a column density on the order of $1 \times 10^{12}$~cm$^{-2}$. Based on observations of B\,0281+357 with the Plateau de Bure interferometer, we estimate an H$_2$ column density of $8 \times 10^{21}$~cm$^{-2}$ \citep{wal16}, which would imply an abundance ratio $^7$LiH/H$_2$ $\sim 10^{-10}$. Since the overall chemical and physical properties of the absorbing gas in the two molecular absorbers toward B\,0218+357 and \PKS1830\ are comparable (see, e.g., \citealt{wal19}), it is difficult to reconcile such several orders of magnitude different relative abundance of LiH between them. It is however likely that these tentative detections were due to spurious artifacts \footnote{Although in principle, we cannot exclude time variations in the absorption profiles.}. Indeed, ALMA archival observations of the same $^7$LiH (1--0) line toward B\,0218+357, taken in October 2018 (project code 2018.1.00187.S) and reduced with a similar process as described in Section~\ref{sec:obs}, show no line absorption feature down to a rms noise level of 1\% of the continuum level, with 1.1~\kms\ velocity resolution (Fig.\,\ref{fig:spec-7LiH-B0218}). For comparison, we also show the spectrum of the $^{13}$CO (4--3) line, observed simultaneously with LiH, which appears to be a similar shape to other lines observed in the past toward this source (i.e., \citealt{wal16}). With a conservative line width of 20~\kms, we derive a column density (3$\sigma$) upper limit for $^7$LiH of $3.5 \times 10^{11}$~cm$^{-2}$, corresponding to an abundance ratio $^7$LiH/H$_2$ $< 5 \times 10^{-11}$, less constraining than the upper limit obtained toward \PKS1830\ by roughly two orders of magnitude.

Our upper limit toward \PKS1830\ is comparable to that obtained by \cite{neu17} toward the Milky Way dense cloud Sgr\,B2, and two orders of magnitude lower than their upper limit along the Galactic diffuse line of sight toward the bright submillimeter continuum source W49N.

In the interstellar medium, most of the lithium is in the ionized form, Li$^+$. The pathways to LiH, either by gas-phase reactions or on dust grain surface, are indeed expected to be slow and inefficient. Compared to the elemental abundance of lithium in the solar system, Li/H $= 1.8 \times 10^{-9}$ \citep{asp09}, our upper limit in the gas in the \PKS1830\ absorber suggests that less than $2\times 10^{-4}$ of the lithium nuclei are bound in LiH.

\section{Discussion} \label{sec:discussion}

Deuterated species had not been detected so far toward \PKS1830, although previous searches focused mainly on DCO$^+$ and DCN (e.g., \citealt{mul06,sha99}). Upper limits on the DCO$^+$/HCO$^+$ and DCN/HCN ratios are $(7-8) \times 10^{-4}$ \citep{mul11} and suggest no strong deuterium fractionation for those two molecules and for the gas column illuminated at that time. However, the column of absorbing gas toward the SW image of \PKS1830\ has a multiphase composition \citep{mul16a,mul16b,mul17,hmul15,sch15}, and changes in the continuum background could occasionally illuminate clouds with different chemical compositions.

The $v=-4$~\kms\ velocity component of the 2019 line of sight shows several peculiar features that are distinguishable from the $v=+5$~\kms\ component:
1) First, the deuterium enhancement in the $v=-4$~\kms\ component appears to be higher than in the $v=+5$~\kms\ component, possibly up to one order of magnitude higher, although the uncertainties are high.
2) The detection of several D-species with a relatively high deuterium enhancement suggests a relatively cold gas temperature, $\lesssim 30$~K for chemical fractionation to be efficient. Previous measurements of the kinetic temperature in the SW line of sight toward \PKS1830\ point toward higher values, $\Tkin \sim 80$~K (\citealt{hen08, mul13}). We note, however, that these measurements were done by assuming a common temperature and density over the whole absorbing gas column and at different epochs.
3) Next, the chemical composition for the $v=-4$~\kms\ component is clearly different from the rest of the absorption system. The good match between the profiles of N$_2$H$^+$ and CH$_3$OH is in agreement with the correlation found by \cite{gra17} in their principal component analysis of Orion clouds. Interestingly, they show that the two species are good tracers of dense cores.
4) The detection of SO$_2$ in the $v=-4$~\kms\ component adds to its chemical peculiarity. SO$_2$ is not seen in diffuse clouds \citep{luc02} but is observed in translucent and thicker clouds (e.g., \citealt{tur95}).
5) The $v=-4$~\kms\ component shows a fast variability, with variations of a factor of two within a couple of months while the $v=+5$~\kms\ component barely changed (Fig.\,\ref{fig:spec-timevar}). This further suggests a relative compactness. Indeed, within this time interval, a plasmon in the quasar's jet, traveling at a superluminal speed of 10$c$ in the plane of the sky, would have an apparent motion covering less than one parsec in the plane of the z=0.89 galaxy. We note that \PKS1830\ was breaking records in $\gamma$-ray and radio flares during spring 2019 (cf. Astronomer's Telegrams \#12594, \#12601, \#12603, \#12622, \#12685, \#12739), and  this activity may be the cause of the absorption profile variability seen in our observations.
6) Looking back at 2009, 2010, and 2011 observations performed by \cite{mul11,mul13}, we note that the centroid velocities of a few species, namely CH$_3$OH, HNCO, CH$_3$NH$_2$, HC$_3$N, and NH$_2$CHO (all complex organic molecules) peak close to $v=-4$~\kms\  and are clearly offset from the average velocity ($v \sim -1$~\kms) of all other ($>20$) molecules detected at the same time.

Overall, these properties suggest that the $v = -4$~ \kms\ component could be associated with a cold dark cloud similar to those in the Milky Way.

\section{Summary and conclusions} \label{sec:conclusions}

We have reported the first detection of deuterated species in the z=0.89 molecular absorber toward the quasar \PKS1830.
Our results are summarized as follows:
\begin{itemize}
\item The deuterated species, namely ND, NH$_2$D, and HDO, are seen enhanced in a narrow (full width at half maximum $\sim 5$~\kms) velocity component along the SW line of sight, with a deuterium fractionation $\sim 100$ times over the primordial D/H. Another weaker velocity component shows deuterium fractionation that is lower by an order of magnitude.
\item Chemically, the component with strong deuterium fractionation is characterized by an enhancement of CH$_3$OH, N$_2$H$^+$, and SO$_2$ with respect to the rest of the absorbing system and is also known to exhibit absorption from several complex organic molecules \citep{mul11}.
\item This velocity component also shows a remarkable time variability with respect to the rest of the absorbing spectrum, with variations by a factor two in optical depth within a couple of months, suggesting a typical size smaller than a parsec.  
\item Based on all these properties, we speculate that this velocity component could be associated with the analog of a Galactic dark cloud. 
\item Finally, the non-detection of $^7$LiH yields a stringent upper limit LiH/H$_2$ $< 4 \times 10^{-13}$ (3$\sigma$) in the interstellar gas of the z=0.89 absorber, suggesting that only a small fraction $\lesssim 10^{-4}$ of the lithium goes into the LiH molecular form.
\end{itemize}
Measurements of deuterium fractionation toward \PKS1830\ could be improved by simultaneous observations of the XD and XH isotopologues, and it will be interesting to monitor the evolution of the absorption profiles and variability of the different velocity components in the future. A comprehensive chemical monitoring would help to further investigate the nature and chemistry of the absorbing clouds toward \PKS1830.

\begin{acknowledgement}
We would like to thank the referee for useful comments.
This paper makes use of the following ALMA data:  ADS/JAO.ALMA\#2018.1.00051.S.,\\
ADS/JAO.ALMA\#2018.1.00692.S,\\
and ADS/JAO.ALMA\#2018.1.00187.S.\\
ALMA is a partnership of ESO (representing its member states), NSF (USA) and NINS (Japan), together with NRC (Canada) and NSC and ASIAA (Taiwan) and KASI (Republic of Korea), in cooperation with the Republic of Chile. The Joint ALMA Observatory is operated by ESO, AUI/NRAO and NAOJ. This research has made use of NASA's Astrophysics Data System.
\end{acknowledgement}

\begin{appendix}

\section{Complementary spectroscopy data}

\subsection{ND}

\begin{table}[ht!]
  \caption{Spectroscopy of the ND ($N$=1--0, $J$=2--1) lines. Adapted from the Cologne Database for Molecular Spectroscopy (\citealt{CDMS01,CDMS05,CDMS16}) and \citep{sai93}.}
\label{tab:spectro-ND21}
\begin{center} \begin{tabular}{ccc}
\hline \hline
$F$'--$F$'' & Frequency & Line \\
            & (GHz)     & strength \\
\hline
 1--1 & 522.001051 & 0.01 \\
 2--3 & 522.008983 & 0.02 \\
 1--2 & 522.010645 & 0.03 \\
 0--1 & 522.010645 & 0.02 \\
 3--2 & 522.018947 & 0.03 \\
 2--1 & 522.025716 & 0.05 \\
 2--1 & 522.028729 & 0.03 \\
 2--2 & 522.035565 & 0.34 \\
 1--1 & 522.035565 & 0.25 \\
 3--3 & 522.035565 & 0.45 \\
 2--2 & 522.044320 & 0.41 \\
 1--2 & 522.045269 & 0.09 \\
 1--0 & 522.045999 & 0.22 \\
 1--1 & 522.047062 & 0.27 \\
 2--3 & 522.052068 & 0.08 \\
 0--1 & 522.056827 & 0.24 \\
 2--1 & 522.061936 & 1.21 \\
 1--2 & 522.062724 & 0.13 \\
 3--2 & 522.071292 & 1.87 \\
 2--1 & 522.071292 & 0.84 \\
 3--2 & 522.077372 & 2.00 \\
 4--3 & 522.077372 & 3.01 \\
 2--1 & 522.079529 & 1.35 \\
 1--1 & 522.080299 & 0.35 \\
 1--0 & 522.080635 & 0.34 \\
 1--1 & 522.081815 & 0.29 \\
 2--2 & 522.087347 & 0.33 \\
 2--2 & 522.089143 & 0.31 \\
 0--1 & 522.090049 & 0.07 \\
 3--3 & 522.094030 & 0.33 \\
 1--2 & 522.097427 & 0.02 \\
 2--1 & 522.104994 & 0.04 \\
 2--3 & 522.105866 & 0.01 \\
 1--1 & 522.115010 & 0.01 \\
 2--1 & 522.125613 & 0.01 \\
 3--2 & 522.129418 & 0.01 \\
\hline
\end{tabular} \end{center}
\end{table}

\begin{table}[ht!]
  \caption{Spectroscopy of the ND ($N$=1--0, $J$=1--1) lines. Adapted from the Cologne Database for Molecular Spectroscopy (\citealt{CDMS01,CDMS05,CDMS16}) and \citep{sai93}.}
\label{tab:spectro-ND11}
\begin{center} \begin{tabular}{ccc}
\hline \hline
$F$'--$F$'' & Frequency & Line \\
            & (GHz)     & strength \\
\hline
 2--1 & 546.102011 & 0.25 \\
 1--1 & 546.102011 & 0.48 \\
 3--2 & 546.111204 & 0.31 \\
 2--2 & 546.111204 & 0.73 \\
 1--2 & 546.111204 & 0.16 \\
 3--3 & 546.127619 & 1.52 \\
 2--3 & 546.128110 & 0.18 \\
 1--1 & 546.128502 & 0.08 \\
 0--1 & 546.128678 & 0.16 \\
 2--2 & 546.137675 & 0.04 \\
 1--2 & 546.138031 & 0.36 \\
 1--1 & 546.140170 & 0.01 \\
 1--0 & 546.147805 & 0.13 \\
 2--1 & 546.147805 & 0.20 \\
 1--1 & 546.147805 & 0.15 \\
 1--2 & 546.149708 & 0.02 \\
 2--3 & 546.154476 & 0.58 \\
 3--2 & 546.163048 & 0.45 \\
 2--2 & 546.163048 & 0.23 \\
 1--2 & 546.163838 & 0.02 \\
 1--0 & 546.173407 & 0.09 \\
 2--1 & 546.174099 & 0.34 \\
 1--1 & 546.174586 & 0.00 \\
 0--1 & 546.174762 & 0.01 \\
 2--1 & 546.181052 & 0.04 \\
 1--1 & 546.181413 & 0.02 \\
 1--0 & 546.185075 & 0.10 \\
 1--1 & 546.186178 & 0.27 \\
 2--2 & 546.189884 & 0.35 \\
 1--2 & 546.190207 & 0.04 \\
 1--2 & 546.201986 & 0.54 \\
 2--1 & 546.207666 & 0.31 \\
 1--1 & 546.207666 & 0.41 \\
 0--1 & 546.207666 & 0.16 \\
 1--1 & 546.219450 & 0.04 \\
 \hline
\end{tabular} \end{center}
\end{table}

\subsection{para-NH$_2$}

\begin{table}[ht!]
\caption{Spectroscopy of the para-NH$_2$ ($N_{Ka},_{Kc}$=$1_{1,0}$--$1_{0,1}$) line. Adapted from the Cologne Database for Molecular Spectroscopy (\citealt{CDMS01,CDMS05,CDMS16}) and \cite{vandis93}.}
\label{tab:spectro-pNH2}
\begin{center} \begin{tabular}{cccc}
\hline \hline
$J$'--$J$'' & $F$'--$F$'' & Frequency & Line \\
       & & (GHz)     & strength \\
\hline
\nicefrac{3}{2}--\nicefrac{1}{2} & \nicefrac{1}{2}--\nicefrac{1}{2} & 461.392564 & 0.15 \\
        & \nicefrac{3}{2}--\nicefrac{1}{2} & 461.399552 & 0.19 \\
        & \nicefrac{1}{2}--\nicefrac{3}{2} & 461.449562 & 0.02 \\
        & \nicefrac{3}{2}--\nicefrac{3}{2} & 461.456550 & 0.14 \\
        & \nicefrac{5}{2}--\nicefrac{3}{2} & 461.465106 & 0.52 \\
\hline
\nicefrac{3}{2}--\nicefrac{3}{2} & \nicefrac{3}{2}--\nicefrac{5}{2} & 462.424981 & 0.39 \\
        & \nicefrac{5}{2}--\nicefrac{5}{2} & 462.433537 & 2.10 \\
        & \nicefrac{1}{2}--\nicefrac{3}{2} & 462.448723 & 0.37 \\
        & \nicefrac{3}{2}--\nicefrac{3}{2} & 462.455711 & 0.91 \\
        & \nicefrac{5}{2}--\nicefrac{3}{2} & 462.464267 & 0.39 \\
        & \nicefrac{1}{2}--\nicefrac{1}{2} & 462.467171 & 0.47 \\
        & \nicefrac{3}{2}--\nicefrac{1}{2} & 462.474158 & 0.37 \\
\hline
\nicefrac{1}{2}--\nicefrac{1}{2} & \nicefrac{1}{2}--\nicefrac{1}{2} & 469.309333 & 0.08 \\
        & \nicefrac{1}{2}--\nicefrac{3}{2} & 469.366331 & 0.60 \\
        & \nicefrac{3}{2}--\nicefrac{1}{2} & 469.383623 & 0.59 \\
        & \nicefrac{3}{2}--\nicefrac{3}{2} & 469.440621 & 0.73 \\
\hline
\nicefrac{1}{2}--\nicefrac{3}{2} & \nicefrac{1}{2}--\nicefrac{3}{2} & 470.365492 & 0.18 \\
        & \nicefrac{1}{2}--\nicefrac{1}{2} & 470.383939 & 0.15 \\
        & \nicefrac{3}{2}--\nicefrac{5}{2} & 470.409052 & 0.50 \\
        & \nicefrac{3}{2}--\nicefrac{3}{2} & 470.439782 & 0.15 \\
        & \nicefrac{3}{2}--\nicefrac{1}{2} & 470.458230 & 0.02 \\
\hline
\end{tabular} \end{center}
\end{table}

\subsection{ortho-NH$_2$D}

\begin{table}[ht!]
\caption{Spectroscopy of the ortho-NH$_2$D ($J_{Ka},_{Kc}$=$1_{1,0}$--$0_{0,0}$) line. Adapted from the Cologne Database for Molecular Spectroscopy (\citealt{CDMS01,CDMS05,CDMS16}).}
\label{tab:spectro-NH2D}
\begin{center} \begin{tabular}{ccc}
\hline \hline
 $F$'--$F$'' & Frequency & Line \\
             & (GHz)     & strength \\
\hline
1--1 & 470.270720 &  3.0 \\
2--1 & 470.271911 &  5.0 \\
0--1 & 470.273657 &  1.0 \\
\hline
\hline
\end{tabular} \end{center}
\end{table}

\end{appendix}
\end{document}